\documentclass[aps,preprint,showpacs]{revtex4}
\usepackage[latin1]{inputenc}
\usepackage{graphicx}
\usepackage{amsmath}
\usepackage{bbm}
\usepackage{subfigure}

\begin{document}
\title{Optimal control of time-dependent targets}
\author{I. Serban, J. Werschnik and E.K.U. Gross}
\affiliation{Institut für Theoretische Physik, Freie Universität  Berlin, Arnimallee 14, 14195 Berlin, Germany}
\date{\today}
\begin{abstract}
In this work, we investigate how and to which extent a quantum system
can be driven along a prescribed path in Hilbert space by a suitably
shaped laser pulse. To calculate the optimal, i.e., the variationally best
pulse, a properly defined functional is maximized. This leads to a
monotonically convergent algorithm which is computationally not more
expensive than the standard optimal-control techniques to push a system,
without specifying the path, from a given initial to a given final state.
The method is successfully applied to drive the time-dependent density
along a given trajectory in real space and to control the time-dependent
occupation numbers of a two-level system and of a one-dimensional model for the
hydrogen atom.
\end{abstract}
\maketitle
\section{Introduction}
Given a quantum-mechanical system, which laser pulse is able to drive the system from state A to state B in a finite time-interval? Which laser pulse maximizes the density in a certain given region in space by the end of the pulse? 
Questions of this kind are addressed by optimal control theory (OCT) in the context of nonrelativistic quantum mechanics.

OCT as a field of mathematics dates back to the late 1950s and is widely applied in engineering. One of the most famous examples in engineering is the reentry problem of a space vehicle into the earth's atmosphere (see, e.g., \cite{SB2}).
The application of OCT to quantum mechanics started in the 1980s \cite{BS90book,HTC83,PDR88,K89}. Due to the enormous progress in the shaping of laser fields \cite{WLPW92}, the control of chemical reactions became within reach. Experiments using closed loop learning (CLL) \cite{JR92} delivered highly convincing results \cite{B97,A98,LMR2001,D2003}.

Calculated pulse shapes may be employed directly in the experimental setup, e.g., as an initial guess for CLL genetic algorithms. Perhaps more important, the theory can be used to decipher the control mechanism embedded in the experimental pulse shapes \cite{B2004}.

The optimal control schemes \cite{TKO92,ZBR98,MT2003} employed so far in theoretical simulations and the experimental applications have been designed to reach a predefined target at the end of a finite time-interval. Little is known about controlling the path the quantum system takes to the desired target, i.e., controlling the trajectory in real space or in quantum number space. To our knowledge, three methods have been proposed so far: A fourth-order Euler-Lagrange equation to determine the envelope of the control field has been derived in Ref. \cite{GGB2002}.  This work, however, is restricted to very simple quantum systems.

Another very elegant method, known as tracking, has been proposed by the authors of Ref. \cite{ZR2003} and Ref. \cite{S2003}. Despite its tremendous success, this method bears an intrinsic difficulty: One has to prescribe a path that is controllable, otherwise singularities in the field appear, because of the one-to-one correspondence between the control field and the given trajectory. In practice, this may require a lot of intuition.

The third method is an optimal control scheme for time-dependent targets \cite{OTR2004}. Basically, it combines optimal control schemes for time-independent targets \cite{ZBR98,ZR98} with an extension to Liouville-space \cite{O2001}. The schemes are generalized by introducing two new parameters like in Ref. \cite{MT2003} and then extended to include time-dependent targets. The new method is monotonically convergent and, in contrast to tracking, does not require a large amount of intuition, i.e., choosing controllable objectives. Furthermore, the method is not restricted to two-level systems. 
While Ref. \cite{OTR2004} presents the monotonically convergent algorithm, the full power of the method has not been exploited as yet. The challenge is the control of a truly time-dependent target represented by a positive-semidefinite, explicitly time-dependent operator. In Sec. II, we describe the general theory along with some examples of such operators. In particular, we discuss the control of occupation numbers in time and the indirect optimization of the dipole operator. The iterative procedure and some numerical details are explained in Sec. III. The results are presented in Sec. IV. 
\section{Theory}
%
We consider an electron in an external potential $V({\bf r})$ under the influence of a laser field. Given an initial state $\Psi({ \bf r },0)=\phi({ \bf r })$, the time evolution of the electron is described by the  time-dependent Schrödinger equation with the laser field modeled in the dipole approximation (length gauge),
\begin{eqnarray}
i\frac{\partial}{\partial t}\Psi({ \bf r },t)&=&\widehat{H}\Psi({ \bf r },t),\label{1SE}\\
\widehat{H}&=&\widehat{H}_0-\hat{{\boldsymbol \mu}}{\boldsymbol \epsilon}(t),\\
\widehat{H}_0&=&\widehat{T}+\widehat{V}
\end{eqnarray}
(atomic units are used throughout: $\hbar = m =e =1 $). Here, $\hat{{\boldsymbol \mu}}=(\hat{\mu}_x,\hat{\mu}_y,\hat{\mu}_z)  $ is the dipole operator and ${\boldsymbol \epsilon}(t)=(\epsilon_x(t),\epsilon_y(t), \epsilon_z(t))$ is the time-dependent electric field. The kinetic energy operator is $\widehat{T}=-\nabla^2/2$.

Our goal is to control the time evolution of the electron by the external field in a way that the time-averaged expectation value of the target operator $\widehat{O}(t)$ is maximized.  Mathematically, this goal corresponds to maximizing the functional
\begin{eqnarray} \label{eq:J1}
J_1[\Psi]&=&\frac{1}{T} \int_0^T \!\! dt \,\, \langle \Psi(t)|\widehat{O}(t)|\Psi(t)\rangle,
\end{eqnarray} 
where $\widehat{O}(t)$ is assumed to be positive-semidefinite.

We want to keep the meaning of the operator $\widehat{O}(t)$ as general as possible at this point. A few examples will be discussed at the end of this section.

Let us define
\begin{eqnarray}
\widehat{O}(t)&=&\widehat{O}_{1}(t)+ 2 T \delta(t-T) \, \widehat{O}_{2},
\end{eqnarray} 
so we can also include targets in our formulation that only depend on the final time $T$ \cite{ZBR98,ZR98,K89}.

The functional $J_1[\Psi]$ will be maximized subject to a number of physical constraints.
The idea is to cast also these constraints in a suitable functional form and then calculate the total variation. Subsequently, we set the total variation to zero and find a set of coupled partial differential equations \cite{K89,PDR88}. The solution of these equations will yield the desired laser field ${ \boldsymbol \epsilon }(t)$.

In more detail, optimizing $J_1$ may possibly lead to fields with very high, or even infinite, total intensity. In order to avoid these strong fields, we include an additional term in the functional which penalizes the total energy of the field,
\begin{eqnarray}
J_2[{ \boldsymbol \epsilon }] &=& - \alpha\int_0^T \!\! dt \,\, { \boldsymbol \epsilon }^2(t). 
\end{eqnarray} 
The penalty factor $\alpha$ is a positive parameter used to weight this part of the functional against the other parts.

The constraint that the electron's wave-function has to fulfill the time-dependent Schrödinger equation is expressed by
\begin{eqnarray}
  J_3[{ \boldsymbol \epsilon },\Psi,\chi]&=&
    - 2 \Im \int_{0}^{T}\!\!  dt \,\, \left\langle\chi(t) \left| \left(i\partial_t
    -\widehat{H}\right) \right| \Psi(t)\right\rangle
\end{eqnarray} 
with a Lagrange multiplier $\chi({ \bf r },t)$. $\Psi({ \bf r },t)$ is the wave function driven by the laser field ${ \boldsymbol \epsilon }(t)$.\\
The Lagrange functional has the form
\begin{eqnarray}
J[\chi,\Psi,{ \boldsymbol \epsilon }] = J_1[\Psi] + J_2[{ \boldsymbol \epsilon }] + J_3[\chi,\Psi,{ \boldsymbol \epsilon }].
\end{eqnarray} 
Setting the variations of the functional with respect to $\chi$, $\Psi$, and ${ \boldsymbol \epsilon }$ independently to zero yields
\begin{eqnarray}
 \alpha \epsilon_j(t) &=& -\Im\langle\chi(t)|\hat{\mu}_j|\Psi(t)\rangle, \label{eq:field} \qquad j=x,y,z\\
0 &=& \left( i \partial_t - \widehat{H} \right) \Psi({ \bf r },t), \label{eq:SE}\\
\Psi({ \bf r },0) &=& \phi({ \bf r }),\label{eq:SE_init}\\
 &&\left(i\partial_t - \widehat{H}\right)\chi({ \bf r },t) + \frac{i}{T}\widehat{O}_{1}(t) \Psi({ \bf r },t)=\nonumber\\
\label{eq:INHSE}
&&i\left(\chi({ \bf r },t)-\widehat{O}_{2}(t) \Psi({ \bf r },t)\right)\delta(t-T).
\end{eqnarray}
Equation (\ref{eq:field}) determines the field from the wave function $\Psi({ \bf r },t)$ and the Lagrange multiplier $\chi({ \bf r },t)$.

Equation (\ref{eq:SE}) is a time-dependent Schrödinger equation for $\Psi({ \bf r },t)$ starting from a given initial state $\phi({ \bf r })$ and driven by the field ${ \boldsymbol \epsilon }(t)$. 
If we require the Lagrange multiplier $\chi({ \bf r },t)$ to be continuous, we can solve the following two equations instead of Eq. (\ref{eq:INHSE}):
\begin{eqnarray}
\label{eq:INHSE2}
 \left(i\partial_t - \widehat{H}\right)\chi({ \bf r },t)&=&-\frac{i}{T}\widehat{O}_{1}(t)\Psi({ \bf r },t),\\
\label{eq:INHSE3}
\chi({ \bf r },T) &=&\widehat{O}_{2} \Psi({ \bf r },T),
\end{eqnarray}
To show this we integrate over Eq. (\ref{eq:INHSE}),
\begin{eqnarray}
\nonumber
&& \lim_{\kappa\to 0}\int_{T-\kappa}^{T+\kappa}\!\!\!\!\!\!dt\left[\left(i\partial_t - \widehat{H}\right)\chi({ \bf r },t) + \frac{i}{T}\widehat{O}_{1}(t) \Psi({ \bf r },t)\right]\\
\label{eq:proof1}
&=&\lim_{\kappa\to 0}\int_{T-\kappa}^{T+\kappa}\!\!\!\!\!\!dt\; i\left(\chi({ \bf r },t)-\widehat{O}_{2}(t) \Psi({ \bf r },t)\right)\delta(t-T).
\end{eqnarray}
The left-hand side of Eq. (\ref{eq:proof1}) is 0 because the integrand is a continuous function. It follows that also the right side must be 0, which implies Eq. (\ref{eq:INHSE3}). From  Eqs. (\ref{eq:INHSE3}) and (\ref{eq:INHSE}) then follows Eq. (\ref{eq:INHSE2}).

Hence, the Lagrange multiplier satisfies an inhomogeneous Schrödinger equation with an initial condition at $t=T$. Its solution can be formally written as
\begin{eqnarray}
\label{eq:SOL_INHSE}
\chi({ \bf r },t)&=&\widehat{U}_{t_0}^{t}\chi({ \bf r },t_0)-
       \frac{1}{T}\int_{t_0}^{t} \!\! d\tau \,\, \widehat{U}_{\tau}^{t}\left(\widehat{O}_{1}(\tau) \, \Psi({ \bf r },\tau)\right),
\end{eqnarray}
where $U_{t_0}^{t}$ is the time-evolution operator defined as  $U_{t_0}^{t}=\mathcal{T}\exp\left(-i \int_{t_0}^{t} \!\! dt' \,\, \widehat{H}(t')\right)$.

The set of equations that we need to solve is now complete: Eqs. (\ref{eq:field}), (\ref{eq:SE}),  (\ref{eq:SE_init}), (\ref{eq:INHSE2}), and (\ref{eq:INHSE3}). To find an optimal field ${ \boldsymbol \epsilon }(t)$ from these equations we use an iterative algorithm which is discussed in the next section.

In principle, we are not restricted to a single particle. The derivation and the algorithm can be generalized to many-particle systems, but except for a few model systems the numerical solution of the many-particle time-dependent Schrödinger equation is not feasible.

We conclude this section with a few examples for the target operator $\widehat{O}(t)$.
\paragraph{Final-time control.}
Since our approach is a generalization of the traditional optimal control formulation given in \cite{K89,ZBR98,ZR98}, we first observe that the latter is trivially recovered as a limiting case by setting
\begin{eqnarray}
\widehat{O}_1(t) = 0, \qquad \widehat{O}_2 = \widehat{P} = | \Phi_f \rangle \langle \Phi_f|.
\end{eqnarray}
Here $\Phi_f$ represents the target final state which the propagated wave function $\Psi({ \bf r },t)$ is supposed to reach at time $T$. 
In this case, the target functional reduces to \cite{K89,ZBR98}
\begin{eqnarray}
J_1 = \langle \Psi(T) | \widehat{P} | \Psi(T) \rangle = | \langle \Psi(T) | \Phi_f \rangle| ^2 .
\end{eqnarray}
The target operator may also be local, as pointed out in  Ref. \cite{ZR98}. If we choose $\widehat{O}_1(t) = 0$ and $\widehat{O}_2 = \delta({ \bf r }-{ \bf r }_0)$ (the density operator), we can maximize the probability density in ${ \bf r }_0$ at $t=T$,
\begin{eqnarray}
\label{eq:loc_op}
J_1 = \int \!\! d{ \bf r } \,\, \langle \Psi(T)| \widehat{O}_2 |\Psi(T) \rangle = n({ \bf r }_0,T). 
\end{eqnarray}
Numerically, the $\delta$ function can be approximated by a sharp Gaussian function. 
\paragraph{Wave-function-follower:}
The most ambitious goal is to find the pulse that forces the system to follow a predefined wave function $\Phi({ \bf r },t)$. If we choose
\begin{eqnarray}
\widehat{O}_{1}(t) &=& |\Phi(t)\rangle\langle\Phi(t)|, \\
\widehat{O}_{2}&=&0,
\end{eqnarray}
the maximization of the time-averaged expectation value of $\widehat{O}_t^{(1)}$ with respect to the field ${ \boldsymbol \epsilon }(t)$ becomes almost equivalent to the inversion of the Schrödinger equation, i.e., for a given function $\Phi({ \bf r },t)$ we find the field $ { \boldsymbol \epsilon }(t)$ 
so that the propagated wave function $\Psi({ \bf r },t)$ comes as close as possible to the target $\Phi({ \bf r },t)$ in the space of admissible control fields.
We can apply this method to the control of time-dependent occupation numbers, if we choose the time-dependent target to be 
\begin{eqnarray}
| \Phi(t) \rangle &=&a_0(t)e^{-i\mathcal{E}_0t}|0\rangle +a_1(t)e^{-i\mathcal{E}_1t}|1\rangle +a_2(t)e^{-i\mathcal{E}_2t}|2\rangle +\ldots\:\:,\\
\hat{H}_0|n\rangle&=&\mathcal{E}_n|n\rangle,\\
\label{eq:td_op}
\widehat{O}_{1}(t)&=& | \Phi(t) \rangle \langle \Phi(t) |.
\end{eqnarray}
The functions $|a_0(t)|^2, |a_1(t)|^2, |a_2(t)|^2, \ldots$ are the predefined time-dependent level occupations which the optimal laser pulse will try to achieve. In general, the functions $a_0(t),a_1(t),a_2(t), \ldots$ can be complex, but as demonstrated in Sec. IV, real functions are sufficient in this case to control the occupations in time. For example, if in a two-level-system the occupation is supposed to oscillate with frequency $\Omega$ we could choose $a_0(t) = \cos(\Omega t)$ and, by normalization,  $a_1(t)=\sin(\Omega t)$. This defines the time-dependent target operator (\ref{eq:td_op}).\\
\paragraph{Moving density.}
The operator used in Eq. (\ref{eq:loc_op}) can be generalized to
\begin{eqnarray}
\widehat{O}_{1}(t)&=&\delta({ \bf r }-{\bf r }_0(t)),\\
J_1 &=& \frac{1}{T} \int_0^T \!\! dt \,\, \langle \Psi(t) |  \delta({ \bf r }-{ \bf r}_0(t)) | \Psi(t) \rangle  \nonumber\\
&=& \frac{1}{T} \int_0^T \!\! dt \,\, n({\bf r}_0(t),t).
\end{eqnarray}
$J_1$ is maximal if the field is able to maximize the density along the trajectory ${\bf r }_0(t)$.

\section{Algorithm and numerical details}
Equipped with the control equations (\ref{eq:field}), (\ref{eq:SE}), and (\ref{eq:INHSE2}) we have to a find an algorithm to solve these equations for ${\boldsymbol \epsilon}(t)$.
In the following, we describe such a scheme which is similar to Ref. \cite{OTR2004}:
\begin{center}
\begin{math}
\begin{array}{l c c l c c c c }
  {\mbox{zero-th step:}} \,\, & \Psi^{(1)}(0) & \overset{{\boldsymbol \epsilon}^{(1)}}{\longrightarrow} &
 \Psi^{(1)}(T) &  &  &  &\\
 {\mbox{kth step:}} \,\, & & & \left[ \Psi^{(k)}(T) \right. & \overset{{\boldsymbol \epsilon}^{(k)}}{\longrightarrow} &
 \left. \Psi^{(k)}(0) \right] & & \\
                   & & & \chi^{(k)}(T) & \overset{\widetilde{{\boldsymbol \epsilon}}^{(k)} 
,\:\:\Psi^{(k)}\:\: }{\longrightarrow} & \chi^{(k)}(0) & &\\
                   & & & & & \left[ \chi^{(k)}(0) \right. & \overset{\widetilde{{\boldsymbol \epsilon}}^{(k)} 
 ,\:\:\Psi^{(k)} \:\: }{\longrightarrow} & \left. \chi^{(k)}(T) \right] \\
                   & & & & & \left[ \Psi^{(k)}(0) \right. & \overset{{\boldsymbol \epsilon}^{(k)}}
{\longrightarrow} & \left. \Psi^{(k)}(T) \right] \\
                   & & & & & \Psi^{(k+1)}(0) & \overset{{\boldsymbol \epsilon}^{(k+1)}}
{\longrightarrow} & \Psi^{(k+1)}(T) .\\
\end{array}
\end{math}
\end{center}
The laser fields used for the propagation are given by
\begin{eqnarray}
\label{feld1}
\widetilde{\epsilon}_j^{(k)}(t) &=& (1-\eta)\epsilon_j^{(k)}(t) 
-  \frac{\eta}{\alpha}\Im\langle\chi^{(k)}(t)|\hat{\mu}_j|\Psi^{(k)}(t)\rangle,\\
\label{feld2}
\epsilon_j^{(k+1)}(t) &=& (1-\gamma)\widetilde{\epsilon}_j^{(k)}(t)
- \frac{\gamma}{\alpha}\Im\langle\chi^{(k)}(t)|\hat{\mu}_j|\Psi^{(k+1)}(t)\rangle \qquad j=x,y,z.
\end{eqnarray}
The initial conditions in every iteration step are
\begin{eqnarray}
  \Psi({\bf r},0)&=& \phi({\bf r}),\\
  \chi({\bf r},T) &=& \widehat{O}_2 \Psi({\bf r},T).
\end{eqnarray}
The propagations in brackets are necessary only if one wants to avoid the storage of the time-dependent wave function and Lagrange multiplier.
Note that the main difference between this iteration and the schemes used in \cite{MT2003} is that one needs to know the time-dependent wave function $\Psi({\bf r},t)$  to solve the inhomogeneous equation (\ref{eq:SOL_INHSE}) for the Lagrange multiplier $\chi({\bf r},t)$. Depending on the operator $\hat{O}_1(t)$, if the inhomogeneity is space- and time-dependent, this may require an additional time propagation.

The choice of $\eta$ and $\gamma$ completes the algorithm. $\gamma = 1$ and $\eta = 1$ correspond to the algorithm suggested in \cite{ZR98}, while the choice $\gamma = 1$ and $\eta = 0$ is analogous to the method used in \cite{K89} with a direct feedback of $\Psi^{(k)}({\bf r},t)$. Further choices are discussed in Refs. \cite{MT2003,OTR2004}.\\
%
%
In the following, we demonstrate the application of our algorithm to two different kinds of time-dependent targets, namely the control of occupation numbers and the control of a local operator.
The first example is a two-level system consisting of states $|0 \rangle$, $|1 \rangle$ with a resonance frequency of $\omega_{01} = \omega_0 - \omega_1 = 0.395$ and the dipole matrix element $P_{01} = \langle 1 | \hat{\mu} |0 \rangle = 1.05$. The second system is a 1D model for hydrogen \cite{SE91}, that has a ``soft'' Coulomb potential,
\begin{equation}
V(x) = - \frac{1}{\sqrt{x^2 + 1}}.
\end{equation}
This type of potential has been used extensively to gain qualitative insights in the behavior of atoms in strong laser pulses \cite{LGE2000,KLEG2001}.

The parameters $\omega_{01}$ and $P_{01}$ of the two-level system are chosen to be identical with the lowest excitation energy and the corresponding dipole matrix element of 1D hydrogen.

The solution of the optimal control Eqs. (\ref{eq:field}), (\ref{eq:SE}), (\ref{eq:INHSE2}), and  (\ref{eq:INHSE3}) requires the integration of the time-dependent Schrödinger equation  with and without inhomogeneity. \\
In the case of the two-level system, one may diagonalize the Hamilton operator 
analytically and therefore calculate the infinitesimal time-evolution operator directly.

The time-dependent Schrödinger equation  for the 1D hydrogen model is solved on a grid, where the infinitesimal time-evolution operator is approximated by the second-order split-operator technique (SPO) \cite{FMF76},
\begin{eqnarray} 
\nonumber
\widehat{U}_{t}^{t+\Delta t}&=&\mathcal{T}\exp\left(-i \int_{t}^{t+\Delta t} \!\! dt' \,\, \widehat{H}(t')\right)\\
\nonumber
\label{eq:spo2nd}
    & \approx  & \exp(-\frac{i}{2}\, \hat {T}\,\Delta t) \exp(-i\, \hat {V}(t)\,\Delta t) \nonumber \\
&  &\exp(-\frac{i}{2}\, \hat {T}\,\Delta t) + O(\Delta t^3).
\end{eqnarray}
For the inhomogeneous Schrödinger equation (\ref{eq:INHSE3}),
the infinitesimal time evolution of $\chi(x,t)$ is given by,
\begin{eqnarray*}
&&\chi(x,t+\Delta t) \nonumber\\
&=&\widehat{U}_{t}^{t+\Delta t}\left(\chi(x,t)-
\frac{1}{T}\int_{t}^{t+\Delta t}\!\!\!\!\!\!\!\!d\tau\,\, 
\widehat{U}_{\tau}^{t}\left(\widehat{O}_1(\tau)\Psi(x,\tau)\right)\right)\\
&\simeq& \widehat{U}_{t}^{t+\Delta t}\left(\chi(x,t)-\Delta t\frac{1}{T} \widehat{O}_1(t)\Psi(x,t)\right),
 \end{eqnarray*}
where we found the above, lowest-order  approximation of the integral to be sufficient.\\
Following the scheme described above, one needs five propagations per iteration step (if we want to avoid storage of the wave function). Within the 2nd order split-operator scheme each time step requires four fast Fourier transforms (FFT) \cite{FFTW98}, because we have to know the wave function and the Lagrange multiplier in real space at every time step to be able to evaluate the field from Eq.  (\ref{eq:field}). This sums up to $2*10^6$ FFTs per $10^5$ time steps and iteration. In comparison, optimal control methods for time-independent targets \cite{MT2003} require four propagations.

Since our hydrogen model can experience ionization, we employ absorbing boundaries to take care of boundary effects (otherwise we will find the outgoing wave incoming from the opposite boundary due to the periodic boundaries introduced by the Fourier transform). The real-space wave function is multiplied with a mask function that falls off like $\cos^{(1/8)}$ at the boundary in every time step.\\
%

\section{Results}
%
\subsubsection{Occupation number control}
%
First, we present the results for the two-level system. The time-dependent target wave function is chosen as $|\Phi(t)\rangle = a_0(t)e^{-i\mathcal{E}_0t}|0\rangle +a_1(t)e^{-i\mathcal{E}_1t}|1\rangle$, where the coefficients  $a_0(t)$ and $a_1(t)$ are real and satisfy $a_0^2(t)+a_1^2(t)=1$. $\hat{O}_1=|\Phi(t)\rangle\langle\Phi(t)|$, $\hat{O}_2=0$.

\begin{figure}[!h]
\centering
     \subfigure[]{
          \label{3eck_2lev_a}
          \includegraphics*[width=.43\textwidth]{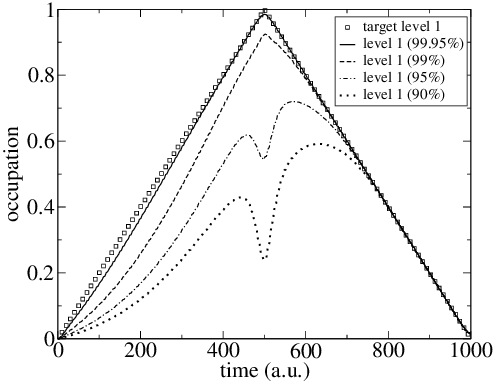}}       
      \hspace{.3in}
      \subfigure[]{
          \label{3eck_2lev_b}
          \includegraphics*[width=.43\textwidth]{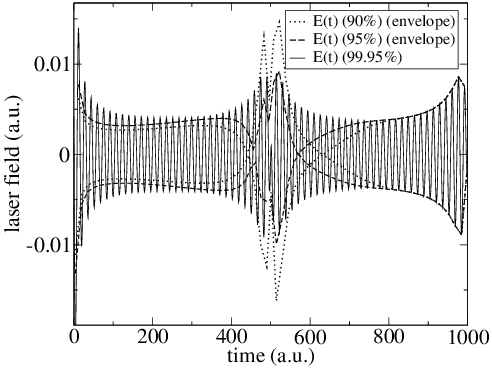}}
      \hspace{.3in}
      \subfigure[]{
          \label{3eck_2lev_c}
          \includegraphics*[width=.43\textwidth]{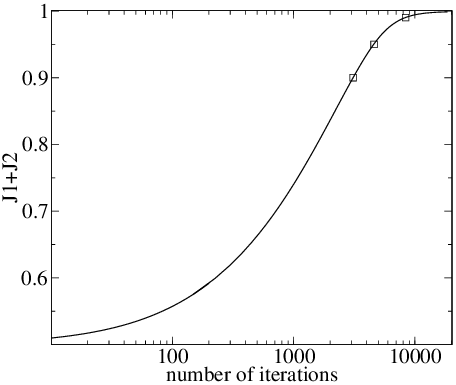} }     
  
        \caption{Target function and calculated occupation numbers for $J_1=0.90,0.95,0.99,0.9995$ (a), optimal field $J_1=0.9995$ and extracted envelopes for $J_1=0.90,0.95$ (b), and the value of the functional $J_1+J_2$ (c).}
\label{3eck_2lev}
\end{figure}

With the parameters $\Delta t = 0.01$, $\alpha$ = 0.05 and the initial guess field  $\epsilon_0(t)$ = $10^{-4}$, the algorithm converges to a final value of $J_1$ = 0.9995 with a difference between two consecutive values of the functional $\delta J^{(n,n+1)}\leq 10^{-8}$ .
Figure \ref{3eck_2lev} shows the numerical results for the time evolution of the occupation numbers [Fig. \ref{3eck_2lev_a}] and the optimized field [Fig. \ref{3eck_2lev_b}]. Figure \ref{3eck_2lev_c} illustrates the monotonic convergence of the functional $J_1+J_2$. The agreement between the calculated occupation and the V-shaped target function, as shown in  Fig. \ref{3eck_2lev_a} is quite remarkable. To illustrate the quality of results associated with different values of $J_1$, we have plotted the occupation curves corresponding to $J_1=0.90,0.95,0.99$ in Fig. \ref{3eck_2lev_a}. Somewhat surprisingly, even if we reach $J_1=0.95$, there is still a sizable difference between the curves. Figure \ref{3eck_2lev_b} shows the envelopes extracted from the laser fields corresponding to $J_1=0.90,0.95$ as well as the optimal field corresponding to  $J_1=0.9995$. The resonance frequency of the system is found within a few steps. Then the algorithm improves the envelope. Figure \ref{3eck_2lev_c} shows the typical convergence behavior: a rapid improvement of the functional in the first few steps, implying that the difference between two consecutive fields is large (\ref{eq:field_diff}), and a slower convergence for the later steps, meaning that the differences between two steps get smaller [see Fig. \ref{3eck_2lev_b}].
%

The same problem was solved for the 1D hydrogen model on a grid. We found $J_1 = 0.97$ with $\delta J^{(n,n+1)} \leq 10^{-5} $. The parameters were 
$\Delta t = 0.005$, $\alpha=1.5$, and the initial choice for the field was again $\epsilon_0(t)$ = $10^{-4}$. From our experience with the two-level system, we expect that the correspondence between the target curve and the optimized curve will not be perfect. 

\begin{figure}[!h]
  \centering
  \subfigure[]{
      \label{3eck_sc_a}
      \includegraphics*[width=.43\textwidth]{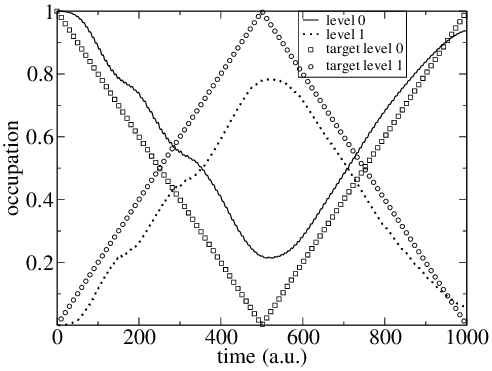}}
      \hspace{.3in}
  \subfigure[]{
      \label{3eck_sc_b}
      \includegraphics*[width=.43\textwidth]{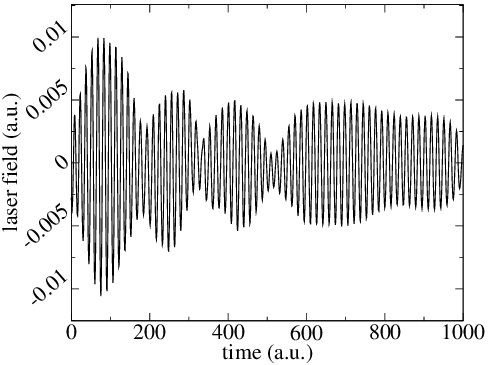}}   
      \hspace{.3in}
  \subfigure[]{
      \label{3eck_sc_c}
      \includegraphics*[width=.43\textwidth]{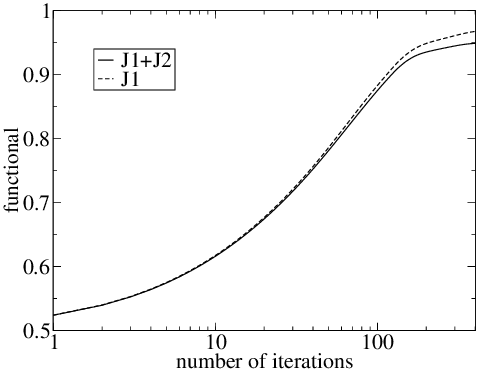}}
      \hspace{.3in}
\caption{Target curves $a_0(t),a_1(t)$ and calculated occupation numbers (a), optimal field (b), and the value of $J_1$ and  $J_1+J_2$ (c).}
\label{3eck_sc}
\end{figure}

\begin{figure}[!h]
  \centering
  \subfigure[]{
      \label{step_2lev_a}
      \includegraphics*[width=.43\textwidth]{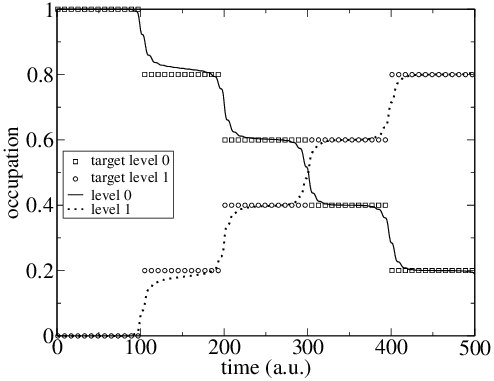} }   
      \hspace{.3in}
   \subfigure[]{
      \label{step_2lev_b}
      \includegraphics*[width=.43\textwidth]{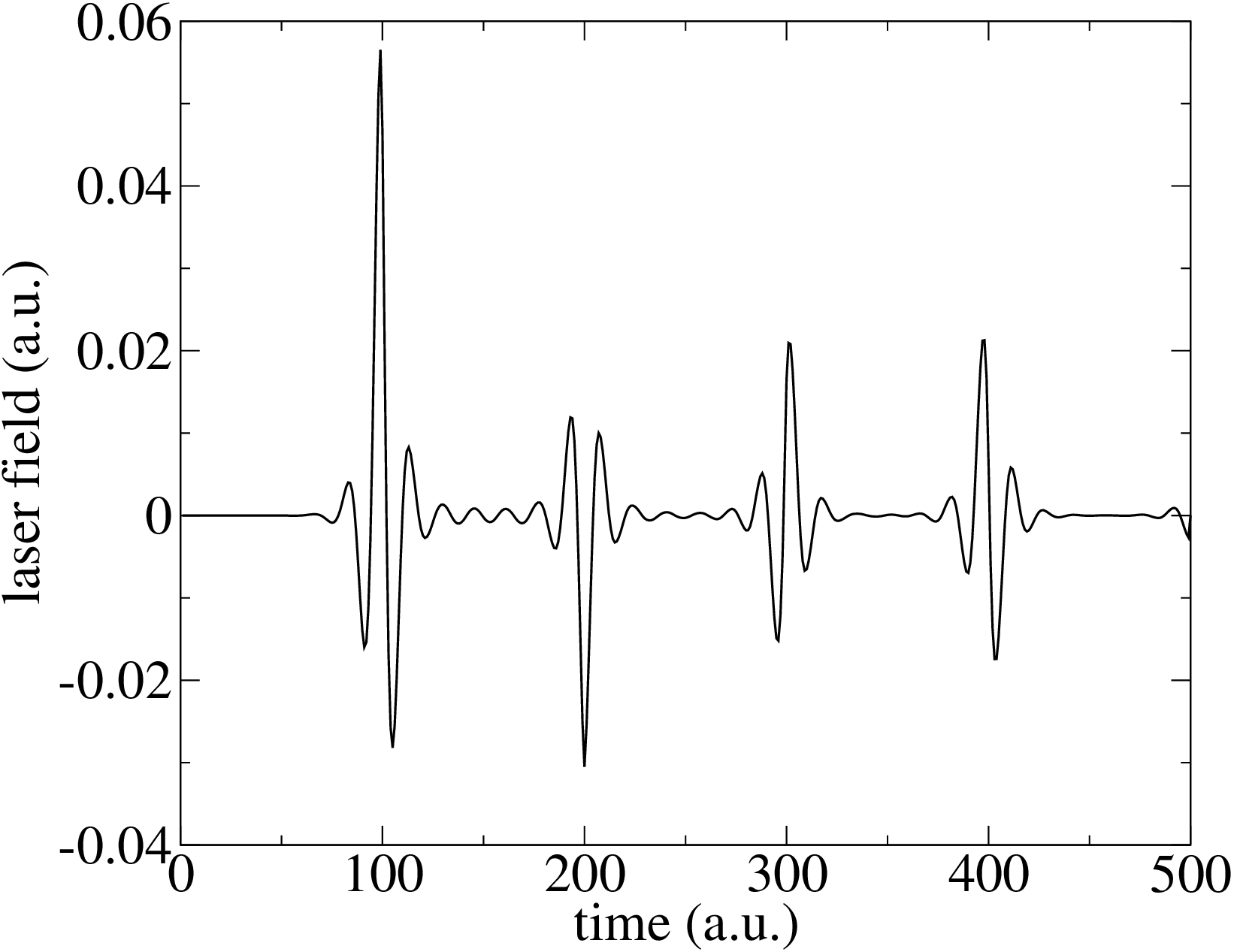}}
      \hspace{.3in}
  \subfigure[]{
      \label{step_2lev_c}
      \includegraphics*[width=.43\textwidth]{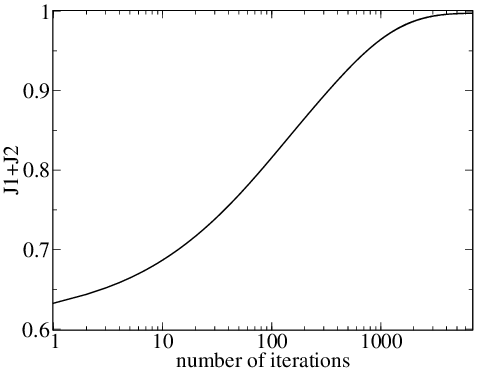}}
      \hspace{.3in}
   \caption{Target curves $b_0(t),b_1(t)$ and calculated occupation numbers (a), optimized field (b), and functional $J_1+J_2$ (c) for $\alpha=0.05$.}
   \label{step_2lev}
\end{figure}

The corresponding numerical results are shown in Fig. \ref{3eck_sc}. Note, that the occupation of higher levels is negligible and that ionization is less than $0.2 \%$. The field is in the weak response regime.\\
%
%
\paragraph*{Tracking versus optimal control.}
Zhu and Rabitz showed \cite{ZR2003} how the exact field necessary to follow a given trajectory can be determined by means of Ehrenfest's theorem. The exact field, however, may have singularities, i.e., the prescribed trajectory is not controllable with a smooth field. If, like in the next example, the target occupation curves $b_0^2(t)$,$b_1^2(t)$  consist of step functions, the exact field must have $\delta$~peaks and, as a consequence, the tracking method cannot produce any useful results. The optimal control approach followed in this paper finds the best compromise between field energy and overlap with the target, yielding reasonable results such as those shown in Fig.~\ref{step_2lev}.
At times where $b_0^2(t)$ becomes discontinuous, the field has intense pulses (see Fig.\ref{step_2lev_b}) consisting of only a few oscillations with the resonance frequency. 

\begin{figure}[!h]
  \centering
  \subfigure[]{
       \label{steps2_a}
       \includegraphics*[width=.43\textwidth]{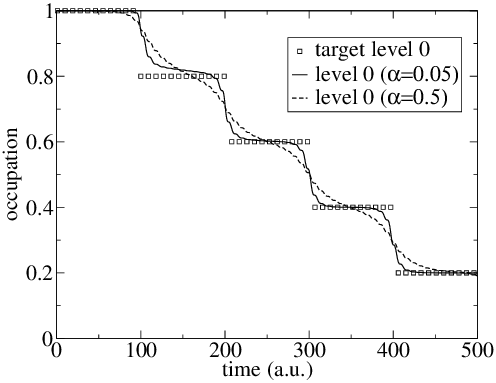}}
       \hspace{.3in}
  \subfigure[]{
       \label{steps2_b}
       \includegraphics*[width=.43\textwidth]{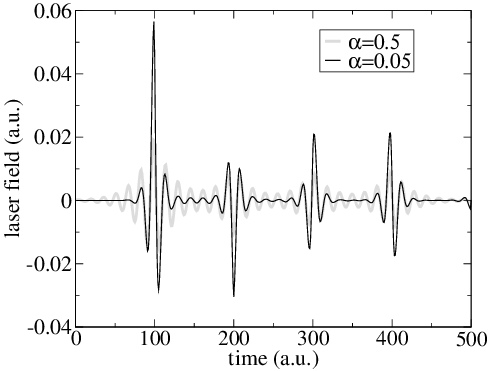} }     
       \hspace{.3in}
   \caption{The influence of the penalty factor $\alpha$: occupation numbers (a) and optimal fields (b).}
   \label{steps2}
\end{figure}

The time-dependent occupation numbers in [Fig.~\ref{step_2lev_a}] deviate slightly from the target curve. They are ``washed out'' at the discontinuity points of the target curve. For larger values of the penalty factor we notice that this broadening of the steps is even more pronounced (see Fig.\ref{steps2_a}). In this case, the width of the pulse envelope becomes broader and the maximum field strength lower [Fig.~\ref{steps2_b}].
%
\subsubsection{Local operator}
%
A very important quantity to control is the time-dependent dipole moment. This quantity, however, cannot be accessed directly because the dipole operator is not positive semidefinite. As an alternative, we choose the time-dependent density operator,
\begin{eqnarray}
\widehat{O}(t)&=& \delta (x-r(t))\label{eq:local_op}\\
\label{eq:delta_approx}
& \approx&\sqrt[4]{\frac{\sigma}{\pi}}e^{-(x-r(t))^2\sigma}.
\end{eqnarray} 
Intuitively, the dipole moment will roughly follow the curve described by $r(t)$ since $r(t)$ governs the movement of the density. 

In the actual computations, we approximate the $\delta$~function by a sharp Gaussian (\ref{eq:delta_approx}).

To test this idea, we first have to choose a reasonable function $r(t)$. For this purpose we solve the time-dependent Schrödinger equation for the 1D hydrogen model with a given laser field $\epsilon(t)$. From the resulting wave function, we calculate the time-dependent expectation value $r(t)= \langle \hat{x} \rangle(t)$. With the function $r(t)$ we then build the target operator (\ref{eq:delta_approx}) and start our optimization with the initial guess $\epsilon_0(t)= 10^{-4}$. \\
%
%
\begin{figure}[!h]
  \centering
  \subfigure[]{
       \label{local1_a}
       \includegraphics*[width=.45\textwidth]{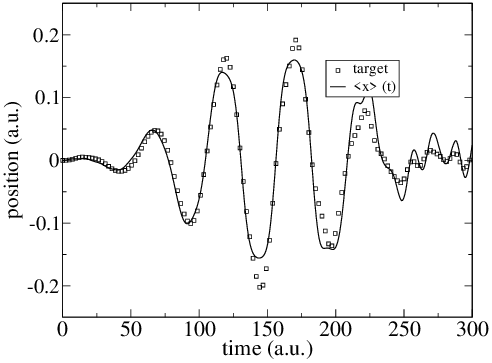}}
       \hspace{.3in}
  \subfigure[]{
       \label{local1_b}
       \includegraphics*[width=.45\textwidth]{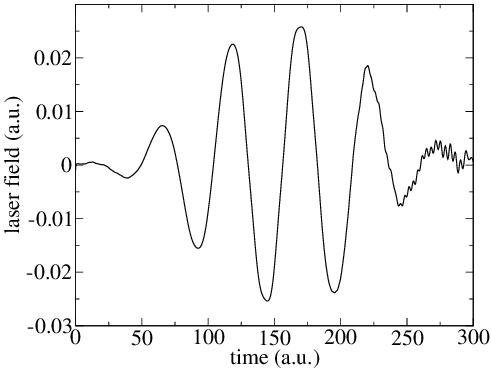}}
       \hspace{.3in}
   \caption{$\langle \hat{x} \rangle$ calculated with target and optimized wave function (a), optimal field (b) (after 1000 iterations) for parameters $\alpha$ = 0.5, $\Delta t = 0.005$, $\epsilon_0=10^{-3}$.}
\label{local1}
\end{figure}

Figure \ref{local1_a} shows that the expectation value $\langle \hat{x} \rangle_{opt}$ calculated with the optimal field $\epsilon_{opt}(t)$ follows the target $r(t)$ rather closely. We do not obtain a perfect correspondence between $\langle \hat{x} \rangle (t)$ and $r(t)$, but the results clearly demonstrate that the algorithm also works for this type of target and, hence,  that the indirect approach to control the dipole moment is appropriate.

\begin{figure}[!h]
   \includegraphics*[width=.45\textwidth]{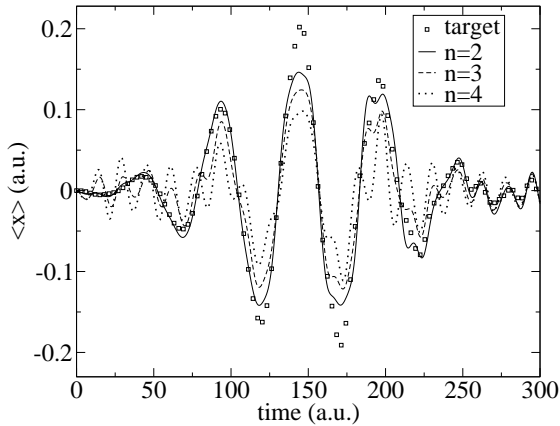}
   \caption{Comparison for the expectation value $\langle \hat{x}\rangle$ with the target trajectory (squares) for different exponents $n$. }
\label{hoch_n_bild}
\end{figure}

As proven in the Appendix, it is also possible to optimize functionals of the type $J_1=\int_0^T dt {\underbrace{\langle\Psi(t)|\widehat{O}(t)|\Psi(t)\rangle}_{I_1}}^n$, $n>1$. Since our integrand $I_1$ is $\leq 1$, the effect will be that $J_1$ carries less weight in the optimization, i.e., the algorithm will try to decrease the field energy. Hence we expect the similarity between the target trajectory $r(t)$ and the calculated expectation value $\langle x\rangle_{opt}$ to be less for increasing $n$. The results for $n=2,3,4$ are shown in Fig.~\ref{hoch_n_bild}. If we build a target functional with the integrand $I_1 \geq 1$ we will find the opposite effect, i.e., $J_1$ will become more important than before. This demonstrates that the parameter $n$ provides a new handle (in addition to the penalty factor $\alpha$) to shift the relative importance of $J_1$ versus $J_2$. 


\section{Conclusion}
%
This work deals with the quantum control of time-dependent targets. In Sec. II, we presented explicit examples of positive-semidefinite target operators designed for the control of time-dependent occupations and of the time-dependent dipole moment. We then applied these operators to control the time evolution of a simple two-level system and of a grid model (1D hydrogen). 
In each case, we find a continuous increase of the value of the functional $J_1 + J_2$. The improvement in the first iteration steps is quite strong, while it takes a large number of iterations to converge the last few percentages. The results also show that a large number of iterations is required to reach perfect agreement with the target trajectories. In the Appendix we prove that by exponentiating the integrand with a positive integer $n>1$, the iteration still converges monotonically. The functional constructed in this way contains two parameters, $\alpha$ and $n$, that allow one to fine-tune the relative importance of $J_1$ and $J_2$. To summarize, we have demonstrated in this work that the quantum control of genuinely time-dependent targets is feasible. In particular, the successful control of the dipole moment in time may open new avenues to optimize high-harmonic generation, which is extremely important for shaping attosecond laser pulses. Work along these lines is in progress.
\begin{acknowledgements}
We would like to thank Stefan Kurth for valuable discussions. This work was supported, in part, by the Deutsche Forschungsgemeinschaft, the EXC!TING Research and Training Network of the European Union and the NANOQUANTA Network of Excellence.
\end{acknowledgements}
\appendix
%
We want to show that the same iteration will converge monotonically also for a functional of the type
\begin{eqnarray}
J_1 & = & \int_0^T\!\!\!\!dt\; \langle\Psi(t)|\widehat{O}(t)|\Psi(t)\rangle^{n},\nonumber
\end{eqnarray}
where $n>1$, $n\in \mathbbm{N}$.
The equation for the Lagrange multiplier then has the form
\begin{eqnarray}
(i\partial_t-H)\chi({\bf r},t) & = &
   -ni\langle\Psi(t)|\widehat{O}(t)|\Psi(t)\rangle^{n-1}
   \widehat{O}(t)\Psi({\bf r},t),\label{hoch_n}\\
\chi({\bf r},T)&=&0.\label{hoch_n_ab}
\end{eqnarray}
Now consider  $a$ and $b$ as real positive numbers. Then
\begin{eqnarray}
a^n-b^n & = & n b^{n-1} (a - b) + \underbrace{a^n + (n - 1) b^n - n b^{n-1} a}_{=A}.
\end{eqnarray}
Next we show that $A=a^n+(n-1)b^n-nb^{n-1}a$ is never negative.
Defining $a = b + \delta$, $\delta\in[-b, \infty)$, we distinguish between two cases.

Case I: $\delta\in[0, \infty)$,
\begin{eqnarray}
&&a^n+(n-1)b^n-nb^{n-1}a\nonumber\\
&& = (b+\delta)^n+(n-1)b^n-nb^{n-1}(b+\delta)  \nonumber\\
&& = b^n + n b^{n-1}\delta + \dots  + n \delta^{n-1}b \nonumber\\
&& +\delta^n+(n-1)b^n-nb^{n-1}\delta\nonumber\\
&& = \frac{n!}{(n-2)!2!}b^{n-2}\delta^2+ \dots  + n \delta^{n-1}b  +\delta^n \geq 0.
\end{eqnarray}

Case II: $\delta\in[-b, 0)$,
\begin{eqnarray}
&&(b + \delta)^n + (n-1)b^n-nb^{n-1}(b+\delta) \nonumber\\ 
&&=\underbrace{b^n}_{\geq0}\Bigg[\left(1+\frac{\delta}{b}\right)^n\nonumber\\
&&+n-1-n-n\frac{\delta}{b}\Bigg].
\end{eqnarray}
To evaluate Case II, we introduce $x= \delta/b$, $x\in[-1,0)$ with,
\begin{eqnarray}
f(x)&=&(1+x)^n-nx-1,\nonumber\\
y&:=&x+1,\nonumber\\
y&\in&[0,1),\nonumber\\
f(y)&=&y^n-ny+n-1,\nonumber\\
f(0)&=&n-1,\nonumber\\
f(1)&=&0,\nonumber\\
f'(y)&=&n\underbrace{(y^{n-1}-1)}_{\leq{0}}\nonumber\\
&\Longrightarrow & f(y)\in[0,n-1]\nonumber\\
&\Longrightarrow & f(y)\geq 0.\nonumber
\end{eqnarray}
Since $f'(y)<0$, the function $f$ must decrease monotonically from $n-1>0$ to 0, so it cannot become negative.
In conclusion,
\begin{eqnarray}
a^n - b^n = nb^{n-1}(a-b)+\underbrace{a^n+(n-1)b^n-nb^{n-1}a}_{A > 0}.\label{positiv}
\end{eqnarray}
The deviation in $J$ between two consecutive steps is,
\begin{eqnarray}
\delta J^{(k+1,k)} & = & J^{(k+1)}-J^{(k)}\nonumber\\
 & = &\int_0^T\!\!\!\!dt\; 
   \Big(\langle\Psi^{(k+1)}(t)|\widehat{O}(t)|\Psi^{(k+1)}(t)\rangle^{n}\nonumber\\
&-&\langle\Psi^{(k)}(t)|\widehat{O}(t)|\Psi^{(k)}(t)\rangle^{n} \nonumber\\
&+& \alpha\left[\epsilon^{(k)}(t)\right]^2-
         \alpha\left[\epsilon^{(k+1)}(t)\right]^2\Big).\nonumber
\end{eqnarray}
If we identify $a(t) = \langle\Psi^{(k+1)}(t)|\widehat{O}(t)|\Psi^{(k+1)}(t)\rangle$ and $b(t) = \langle\Psi^{(k)}(t)|\widehat{O}(t)|\Psi^{(k)}(t)\rangle$ and use Eq. (\ref{positiv}), 
\begin{eqnarray}
\delta J^{(k+1,k)}& = & \int_0^T\!\!\!\!dt\;\Bigg( A(t) +
       \alpha\left[\epsilon^{(k)}(t)\right]^2-
       \alpha\left[\epsilon^{(k+1)}(t)\right]^2  \nonumber\\
& + & n \langle\Psi^{(k)}(t)|\widehat{O}(t)|\Psi^{(k)}(t)\rangle^{n-1} \nonumber\\
& & \left(\langle\Psi^{(k+1)}(t)|\widehat{O}(t)|\Psi^{(k+1)}(t)\rangle-
            \langle\Psi^{(k)}(t)|\widehat{O}(t)|\Psi^{(k)}(t)\rangle\right)\Bigg),\nonumber
\end{eqnarray}
where we have separated the positive term $A(t)=a^n(t)+(n-1)\,b^n(t)-nb^{n-1}(t)\,a(t)$. We define $B(t)=n \langle\Psi^{(k)}(t)|\widehat{O}(t)|\Psi^{(k)}(t)\rangle^{n-1}\langle\delta\Psi^{(k+1,k)}(t)|\widehat{O}(t)|\delta\Psi^{(k+1,k)}(t)\rangle\geq0$ and rewrite $\delta J^{(k+1,k)}$ as 
\begin{eqnarray}
\delta J^{(k+1,k)}& = &\int_0^T\!\!\!\!dt\; \Big(A(t) + B(t)+
       \alpha\left[\epsilon^{(k)}(t)\right]^2-
       \alpha\left[\epsilon^{(k+1)}(t)\right]^2\nonumber\\
& + & n \langle\Psi^{(k)}(t)|\widehat{O}(t)|\Psi^{(k)}(t)\rangle^{n-1}
    2\Re\langle\Psi^{(k)}(t)|\widehat{O}(t)|\delta\Psi^{(k+1,k)}(t)\rangle\Big),\nonumber
\end{eqnarray}
where $\delta\Psi^{(k+1,k)}({\bf r},t)=\Psi^{(k+1)}({\bf r},t) - \Psi^{(k)}({\bf r},t)$. We use the equation for the Lagrange multiplier (\ref{hoch_n}) and obtain
\begin{eqnarray}
\delta J^{(k+1,k)}& = & \int_0^T\!\!\!\!dt\; \Big(A(t) + B(t)\nonumber\\
& + &    \alpha\left[\epsilon^{(k)}(t)\right]^2-
       \alpha\left[\epsilon^{(k+1)}(t)\right]^2  \nonumber\\
& + & 2\Re \left\langle-\left(\partial_t+i\widetilde{H}^{(k)}\right)\chi^{(k)}(t)|\delta\Psi^{(k+1,k)}(t)\right\rangle\Big),
     \label{conect2}
\end{eqnarray}
where $\widetilde{H}^{(k)}=\widehat{H}_0-\hat{\boldsymbol \mu}\widetilde{\epsilon}^{(k)}$,
\begin{eqnarray}
\delta J^{(k+1,k)}& = & \int_0^T\!\!\!\!dt\;\Big(A(t) + B(t) \nonumber\\ 
& + &   \alpha\left[\epsilon^{(k)}(t)\right]^2-
       \alpha\left[\epsilon^{(k+1)}(t)\right]^2 \nonumber\\
& + & 2\Im\left\langle\chi^{(k)}(t)|\left(i\partial_t-\widetilde{H}^{(k)}\right)|\delta\Psi^{(k+1,k)}(t)\right\rangle\Big) \nonumber\\
  &+&  \underbrace{2\Re\langle\chi(t)|\delta\Psi^{(k+1,k)}(t)\rangle|_0^T}_{0}.
\label{hoch_n_conect}
\end{eqnarray}
For the last term in Eq.~(\ref{hoch_n_conect}) we used the fact that $\delta\Psi^{(k+1,k)}({\bf r},0)$ = 0 since the initial state for the wave function is fixed and $\chi({\bf r},T)$ = 0 because of Eq.~(\ref{hoch_n_ab}).

We use the time-dependent Schrödinger equation (\ref{eq:SE}) for $\Psi^{(k)}$ and $\Psi^{(k+1)}$, where $\widehat{H}^{(k)}=\widehat{H}_0-\hat{\boldsymbol\mu}{\epsilon}^{(k)}$,
\begin{eqnarray}
&&\left(i\partial_t-\widetilde{H}^{(k)} \right) \delta \Psi^{(k+1,k)}({\bf r},t) \nonumber\\
&=&\left(i\partial_t-\widetilde{H}^{(k)} \right) \left( \Psi^{(k+1)}({\bf r},t)-\Psi^{(k)}({\bf r},t) \right)\nonumber\\
&=&\left( \widehat{H}^{(k+1)}-\widetilde{H}^{(k)} \right) \Psi^{(k+1)}({\bf r},t)  \nonumber\\
\nonumber
  &&-\left( \widehat{H}^{(k)} - \widetilde{H}^{(k)} \right) \Psi^{(k)}({\bf r},t) \\
\nonumber
&=&-\left( {\boldsymbol \epsilon}^{(k+1)}(t) - \widetilde{{\boldsymbol \epsilon}}^{(k)}(t) \right) \hat{{\boldsymbol \mu}}({\bf r}) \Psi^{(k+1)}({\bf r},t) \\
&&+\left({\boldsymbol \epsilon}^{(k)}(t)-\widetilde{{\boldsymbol \epsilon}}^{(k)}(t)\right) \hat{{\boldsymbol \mu}}({\bf r})\Psi^{(k)}({\bf r},t).
\end{eqnarray}
Consequently, the change in the Lagrange functional becomes
\begin{eqnarray}
\nonumber
\delta J^{(k+1,k)} & = & \int_0^T  \!\! dt \,\, 
\bigg( A(t) + B(t)+   \langle \delta \Psi^{(k+1,k)}(t) | \widehat{O}(t) | \delta \Psi^{(k+1,k)}(t) \rangle \\
\nonumber
&&- \alpha \left( \left[ {\boldsymbol \epsilon}^{(k+1)} \right]^2- \left[ {\boldsymbol \epsilon}^{(k)}(t) \right]^2 \right) \\
\nonumber
&&- 2 \Im \langle \chi^{(k)}(t)| \hat{{\boldsymbol \mu}} |\Psi^{(k+1)}(t) \rangle ( {\boldsymbol \epsilon}^{(k+1)}(t)- \widetilde{{\boldsymbol \epsilon}}^{(k)}(t))\\
&&+ 2 \Im \langle \chi^{(k)}(t)| \hat{{\boldsymbol \mu}} |\Psi^{(k)}(t) \rangle ( {\boldsymbol \epsilon}^{(k)}(t) - \widetilde{{\boldsymbol \epsilon}}^{(k)}(t)) \bigg).
\end{eqnarray}
Finally, using Eqs.~(\ref{feld1}) and (\ref{feld2}), we find
\begin{eqnarray}
\nonumber
&&\delta J^{(k+1,k)} \nonumber\\
&=& \int_0^T \!\!  dt \, \, 
       \bigg( A(t) + B(t)+ \langle\delta\Psi^{(k+1,k)}(t)|\widehat{O}(t)|\delta\Psi^{(k+1,k)}(t) \rangle \nonumber\\
\nonumber
      & - &\alpha \left[ {\boldsymbol \epsilon}^{(k+1)}(t) \right]^2 + \alpha \left[ {\boldsymbol \epsilon}^{(k)} \right]^2   \\
       & + & 2\bigg({\boldsymbol \epsilon}^{(k+1)}(t)-(1-\gamma) \widetilde{{\boldsymbol \epsilon}}^{(k)}(t)\bigg) \frac{\alpha}{\gamma}
\nonumber
             \bigg({\boldsymbol \epsilon}^{(k+1)}(t)-\widetilde{{\boldsymbol \epsilon}}^{(k)}(t)\bigg)\\
       & - & 2\bigg(\widetilde{{\boldsymbol \epsilon}}^{(k)}(t)-(1-\eta){\boldsymbol \epsilon}^{(k)}(t)\bigg) \frac{\alpha}{\eta}
         \bigg({\boldsymbol \epsilon}^{(k)}(t)-\widetilde{{\boldsymbol \epsilon}}^{(k)}(t)\bigg)\bigg)\\
\nonumber
& = &\int_0^T\!\! dt \,\,   \Bigg( A(t)+B(t) + \langle \delta \Psi^{(k+1,k)}(t)|\widehat{O}(t)|\delta\Psi^{(k+1,k)}(t) \rangle \\
\nonumber
&+& \alpha \left(\frac{2}{\gamma}-1\right) \left( {\boldsymbol \epsilon}^{(k+1)}(t) - \widetilde{{\boldsymbol \epsilon}}^{(k)}(t) \right)^2 \\
\label{eq:field_diff}
&+& \alpha \left(\frac{2}{\eta}-1\right) \left( {\boldsymbol \epsilon}^{(k)}(t) - \widetilde{{\boldsymbol \epsilon}}^{(k)}(t) \right)^2\Bigg).
\end{eqnarray}
For $\eta, \gamma \in [0,2]$ (similar to \cite{MT2003} and \cite{OTR2004}), the iteration converges monotonically, i.e., $\delta J^{(k+1,k)}\geq 0$.
This iteration converges  monotonically and quadratically in terms of the field deviations between two iterations. 

We emphasize that this proof is true only if the solution of the time-dependent (in)homogeneous Schrödinger equation is exact in each step. Numerical implementations are of course always approximate and, as a consequence, it may happen that the value of the functional $J$ decreases. This, on the other hand provides a test of the accuracy of the propagation method.\\


\bibliographystyle{prsty}
\bibliography{paper}
\end{document}